\documentclass[prl,twocolumn]{revtex4}
\usepackage{epsfig}
\usepackage{bm}
\usepackage{amsmath}

\begin{document}

\title{Lagrangian chaos and turbulent diffusivity}

\author{A. Bershadskii}

\affiliation{
ICAR, P.O. Box 31155, Jerusalem 91000, Israel
}

\begin{abstract}
  Passive scalar mixing, produced by Lagrangian chaos generated a) by quasi-periodic (integrable) motion of three quasi-point vortices and b) by chaotic motion of three and six quasi-point vortices, has been studied and compared with turbulent mixing of passive scalar in 2D and 3D steady isotropic homogeneous turbulence and in turbulent wakes behind grid and behind cylinder. Results of numerical and laboratory experiments have been used and effective diffusivity approximation as well as distributed chaos approach have been applied to this problem. 
\end{abstract}

\maketitle

\section{Introduction}

  Let us start from the Navier-Stokes equations for incompressible fluids \cite{my}

$$  
  \partial_t {\bf v} + ({\bf v}\cdot \nabla) {\bf v} = -\nabla p +\nu \nabla^2 {\bf v}+\mathbf{f}_{\bf v}, \eqno{(1)}
$$
$$  
  \nabla \cdot \bf{v}=0   \eqno{(2)}
$$    

Energy dissipation rate  
$$
\varepsilon = \nu~ \Omega,  \eqno{(3)}
$$ 
where the enstrophy
$$
\Omega = \langle {\boldsymbol \omega} ({\bf x},t)^2 \rangle_{{\bf x}},   \eqno{(4)}
$$
${\boldsymbol \omega} ({\bf x},t)= \nabla \times {\bf v}$ is vorticity, $\nu$ is viscosity and $\langle ... \rangle_{{\bf x}}$ means spatial average.\\

  The Kolmogorov-Obukhov phenomenology assumes existence of an inertial range of scales where viscosity $\nu$ plays its role through the energy dissipation rate Eq. (3) only and the kinetic energy spectrum \cite{my}
$$ 
E_v(k) \propto \varepsilon^{2/3} k^{-5/3}  \eqno{(5)}
$$
In this phenomenology the enstrophy $\Omega$ is assumed to be approximately constant for the inertial range of scales. It is also assumed that in the limit $\nu \rightarrow 0$ the ensrophy $\Omega \rightarrow \infty$ so that the energy dissipation rate $\varepsilon$ is finite in this limit.\\

  An example of a flow with infinite enstrophy at $\nu = 0$ can be found in two-dimensional inviscid case. In this case Eqs. (1-2) allows a superposition of point vortices as a solution. The vorticity field 
$$ 
{\boldsymbol \omega} ({\bf x},t) = \sum_i \Gamma_i \delta ({\bf r} -{\bf r}_i)  \eqno{(6)}
$$ 
where $\Gamma_i$ (circulations) are some constants and ${\bf r_i}$ are positions of the point vortices. Motion of the system of the point vortices on the unbounded plane is described by a Hamiltonian system of equations
$$
\Gamma_i \dot{x_i} = \frac{\partial H}{\partial y_i}, ~~~~~~~~~~  \Gamma_i \dot{y_i} = -\frac{\partial H}{\partial x_i} 
  \eqno{(7)}
$$
where the Hamiltonian
$$
H = -\frac{1}{4\pi} \sum_{i<j}^N \Gamma_i\Gamma_j \ln [(x_i-x_j)^2 +(y_i-y_j)^2]   \eqno{(8)}
$$
$N$ is the number of the point vortices (see, for instance, Ref. \cite{aref}. This example, with certain modification, will be useful for further consideration.\\

  Mixing of passive scalar $\theta ({\bf x}, t)$ by a velocity field ${\bf v} ({\bf x}, t)$ can be described by equation
$$
\partial_t \theta + (\bf{v}\cdot\nabla)\theta= \kappa\nabla^2\theta+f_{\theta}  \eqno{(9)}
$$  
The Obukhov-Corrsin phenomenology for the passive scalar mixing is an extension of the Kolmogorov-Obukhov phenomenology on the passive scalar mixing. It is assumed that there exists an inertial-convection range of scales where the molecular diffusivity $\kappa$ and viscosity play their role through the combination \cite{my}
$$
\chi \varepsilon^{-1/3} = 2\kappa \langle (\nabla \theta)^2 \rangle_{\bf x} (\nu \Omega))^{-1/3} 
$$
only (the $\chi= 2\kappa \langle (\nabla \theta)^2 \rangle_{\bf x}$ is an analogue of the dissipation rate for the passive scalar), and power spectrum of the passive scalar fluctuations 
$$
E(k) \propto   \kappa \langle (\nabla \theta)^2 \rangle_{\bf x} (\nu \Omega)^{-1/3} k^{-5/3}  \eqno{(10)}
$$

\section{Mixing by quasi-periodic motion of three quasi-point vortices} 

 Results of a numerical experiment with mixing of the passive scalar by three quasi-point vortices in a circular cylinder were reported in Ref. \cite{yuan}. In order to take into account the impenetrable boundary conditions ($({\bf v}\cdot{\bf n})=0$ along the boundary) the authors added to the system of the three real vortices three image vortices of strength $- \Gamma_i$ and located at the mirror image of ${\bf r}_i$. The Hamiltonian was changed accordingly.
 
 In this numerical simulations dynamics of the quasi-point vortices was described by the Hamiltonian corresponding to the point vortices approach, whereas the point vortices were replaced by tiny vortex patches where the vorticity was uniformly distributed. One can expect (and it was confirmed in the numerical experiment) that this approach is valid  as long as the quasi-point vortices do not come too close to each other. There are no singularities in this motion (cf. Eq. (6) for the point vortices) and the enstrophy $\Omega$ is finite and constant. \\
 
   Figure 1 (adapted from Fig. 4a of the Ref. \cite{yuan}) shows a regular trajectory observed in this experiment for initial conditions ${\bf r}_1(0) = (0.1,0.7)$, ${\bf r}_2(0) = (0.5,0.5)$, and ${\bf r}_3(0) = (0.5,0.3)$, with $\Gamma_1=\Gamma_2=\Gamma_3$=constant. The motion of vortices is quasi-periodic (integrable) in this case. Then the velocity field, generated by these quasi-point vortices, was substituted into the passive scalar equation Eq. (9) with a scalar source continually injecting passive scalar density at low values of the wavenumbers $k$ at a constant rate.\\

 Despite the quasi-periodic (integrable) motion of the quasi-point vortices a Lagrangian chaos was generated by this motion and power spectrum of the passive scalar concentration fluctuations was a continuous one. This spectrum is shown in Fig. 2 in the log-log scales (the spectral data were taken from Fig. 7 of the Ref. \cite{yuan}).

\section{Effective diffusivity}

  In order to indicate existence of a power-law range of scales (two decades long) in the spectrum a straight line with slope '-5/3' in the log-log scales has been drawn in the Fig. 2. There is no viscosity in this numerical experiment. Therefore one cannot apply  the Obukhov-Corrsin phenomenology Eq. (10). The inverse cascade phenomenology (see, for instance, Ref. \cite{chen} and references therein) also cannot be expected here.\\ 
\begin{figure} \vspace{-0.6cm}\centering
\epsfig{width=.37\textwidth,file=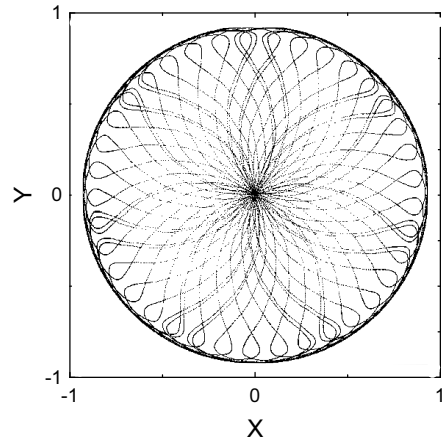} \vspace{-2.1cm}
\caption{A regular trajectory observed in the quasi-periodic motion of the three quasi-point vortices.} 
\end{figure}
\begin{figure} \vspace{-0.5cm}\centering
\epsfig{width=.45 \textwidth,file=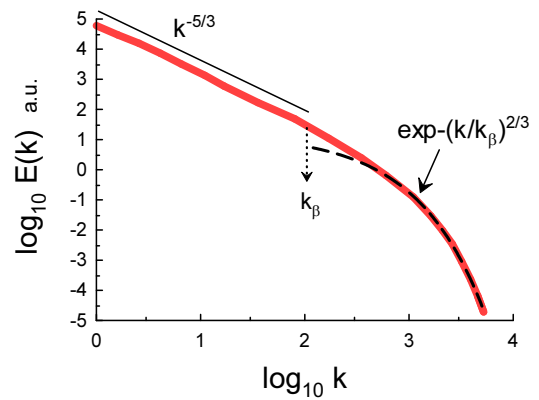} \vspace{-3.8cm}
\caption{Power spectrum of the passive scalar fluctuations corresponding to the Lagrangian chaos generated by the quasi-periodic motion of the three quasi-point vortices.} 
\end{figure}
  
  The notion of effective diffusivity $D$ is a long known one used both in the numerical simulations and in the theory (see, for instance, Refs. \cite{my},\cite{bof} and references therein). Let us generalize the Obukhov-Corrsin phenomenology and replace the viscosity $\nu$ in the Eq. (10) by the effective diffusivity $D$
$$
E(k) \propto   \kappa \langle (\nabla \theta)^2 \rangle_{\bf x} (D \Omega)^{-1/3} k^{-5/3}  \eqno{(11)}
$$
  
  Since the Hamiltonian of the system is an invariant of the motion
$$
\frac{dH}{dt} = 0 \eqno{(12)}
$$
one can write for this system
$$
D = c~ |H|^{1/2}  \eqno{(13)}
$$
where $c$ is a dimensionless constant.  On the other hand, one can estimate the effective diffusivity $D$ as
$$
D=v_cl_c   \eqno{(14)}
$$
where $l_c$ and $v_c$ are the characteristic spatial scale and velocity respectively. If one uses the wavenumber terms $l_c \propto 1/k_c$, then 
$$
D \propto v_ck_c^{-1}  \eqno{(15)}
$$
or
$$
v_c \propto D~k_c  \eqno{(16)}
$$
  
  In order to understand what happens for the spatial scales which are smaller then those belonging to the power-law range let the characteristic velocity $v_c$ and characteristic spatial scale $l_c$ (or wavenumber $k_c$) fluctuate remaining the effective diffusivity the same constant, and let us use the distributed chaos approach for the fluctuating $k_c$ \cite{b1}: 
$$
E(k) \propto \int_0^{\infty} P(k_c) \exp -(k/k_c)dk_c  \propto \exp-(k/k_{\beta})^{\beta}  \eqno{(17)}
$$    
where $P(k_c)$ is distribution of the fluctuating $k_c$. 

 Let us estimate asymptote of $P(k_c)$ at $k_c \rightarrow \infty$ from the Eq. (17) 
$$
P(k_c) \propto k_c^{-1 + \beta/[2(1-\beta)]}~\exp(-bk_c^{\beta/(1-\beta)}) \eqno{(18)}
$$
$b$ is a constant \cite{jon}. If fluctuating $v_c$ has Gaussian distribution, then from the Eq. (16) one obtains Gaussian distribution for the fluctuating $k_c$ as well. The distribution Eq. (18) is Gaussian when $\beta = 2/3$ and one obtains
$$
E(k) \propto \exp-(k/k_{\beta})^{2/3}  \eqno{(19)}
$$
   The dashed curve in the Fig. 2 indicates correspondence to the spectrum Eq. (19) and the dotted arrow indicates position of the scale $k_{\beta}$. Actually the scale $k_{\beta}$ separates the power-law ('-5/3') and the distributed chaos regions of scales (see also below).

\section{Mixing in three-dimensional isotropic homogeneous turbulence}

  Despite the fact that the spectrum shown in the Fig. 2 corresponds to the passive scalar mixing produced by quasi-periodic motion of the three quasi-point vortices (rather minimal system and motion type) it is interesting to compare the Fig. 2 with corresponding spectra observed in the laboratory and direct numerical simulations of three-dimensional isotropic homogeneous turbulence (especially taking into account the role of the vortex filaments in three-dimensional turbulence for moderate and large values of Reynolds number, see Ref. \cite{yzs},\cite{isy} and references therein). \\
  
\begin{figure} \vspace{-1.8cm}\centering
\epsfig{width=.45\textwidth,file=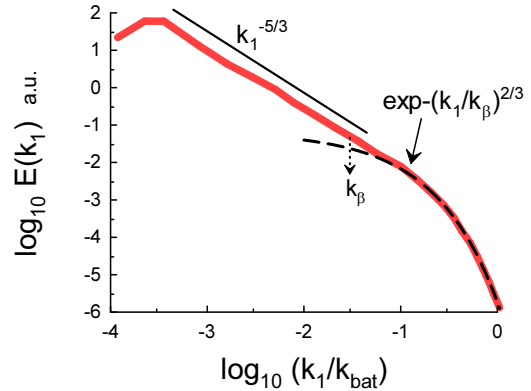} \vspace{-3.8cm}
\caption{Power spectrum of the passive scalar fluctuations generated by a quasi-isotropic grid turbulence in a laboratory experiment at $Re_{\lambda}=582$ and $Sc\simeq 0.71$. } 
\end{figure}
\begin{figure} \vspace{-0.5cm}\centering
\epsfig{width=.45\textwidth,file=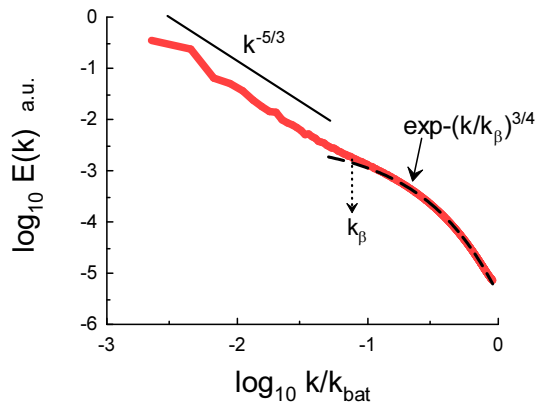} \vspace{-3.6cm}
\caption{Power spectrum of the passive scalar fluctuations generated by a steady isotropic homogeneous turbulence at $Re_{\lambda}=427$ and $Sc=1$. } 
\end{figure}
  
  Wake behind a grid is often used for laboratory simulations of the three-dimensional isotropic homogeneous turbulence \cite{my}. Figure 3 shows power spectrum of the passive scalar fluctuations (temperature in this case) for the Taylor-Reynolds number $Re_{\lambda}=582$ and the Schmidt number $Sc=\nu/\kappa \simeq 0.71$ observed in a laboratory experiment in turbulent wake behind a grid \cite{myd}. The passive scalar was induced in the flow by imposed mean temperature gradient. The spectral data for the Fig. 3 were taken from Fig. 4 of the Ref. \cite{myd} ($k_1$ - the longitudinal wavenumber, has been normalized by the Batchelor scale $k_{bat} = [\varepsilon/(\nu \kappa^2)]^{1/4}$ \cite{my}). As in the Fig. 2  the straight line with slope '-5/3' in the log-log scales has been drawn in the Fig. 3 for reference. The dashed curve in the Fig. 3 indicates correspondence to the stretched exponential spectrum Eq. (19). The dotted arrow indicates position of the normalized scale $k_{\beta}$. The scale $k_{\beta}$ separates the power-law ('-5/3') and the distributed chaos regions of scales, as in the Fig. 2.\\
  
    The effective diffusivity in this case can be based on the Chkhetiani invariant (velocity-vorticity correlation integral \cite{otto1}-\cite{b2}):
$$
I =     \int \langle {\bf v} ({\bf x},t) \cdot  {\boldsymbol \omega} ({\bf x} + {\bf r}, t) \rangle d{\bf r}
$$
as
$$
D = C  \lvert I \rvert^{1/2}  \eqno{(20)}
$$  
(where $C$ is a dimensionless constant), instead of the Hamiltonian $H$ used in the point-vortices case  Eq. (13). In the two-dimensional case the Chkhetiani invariant is equal to zero and, consequently, cannot be used for this purpose. It should be noted the Chkhetiani invariant is related to helicity: $h={\bf v} \cdot {\boldsymbol \omega}$ \cite{otto1} and, through this relation, to the Lagrangian relabeling symmetry \cite{y}.\\

  Figure 4 shows power spectrum of the passive scalar fluctuations for $Re_{\lambda}=427$ and $Sc=1$ observed in a direct numerical simulation of the steady isotropic homogeneous turbulence \cite{wg}. The $\mathbf{f}_{\bf v}$ (Gaussian random force, Eq. (1)) and $f_{\theta}$ (scalar source, Eq. (9)) are delta-correlated in time and (as for the three vortices numerical experiment) have been added in the low-wavenumbers range. As in the Fig. 2  the straight line with slope '-5/3' in the log-log scales has been drawn in the Fig. 4 for reference.
      
      The dashed curve in the Fig. 4 indicates correspondence to the stretched exponential spectrum Eq. (17) with $\beta =3/4$, that is related to the Birkhoff-Saffman invariant of the Navier-Stokes Eqs. (1-2) \cite{b1}. This invariant corresponds to the linear momentum conservation and, according to the Noether's theorem \cite{js}, to the spatial translational symmetry of the Eqs. (1-2) \cite{b1},\cite{saf},\cite{d}. \\
      
      The value $\beta =3/4$ is rather close to the value $\beta =2/3$. An interplay (renormalization) between the Birkhoff-Saffman invariant (velocity-velocity correlation integral \cite{d}) and the Chkhetiani invariant (velocity-vorticity correlation integral \cite{otto1}-\cite{b2}) in the three-dimensional distributed chaos can be rather interesting. If we apply the effective diffusivity approximation to the spectral data shown in the Fig. 4 (i.e. an approximation by the stretched exponential with $\beta =2/3$ instead of the $\beta =3/4$), then we obtain value of $k_{\beta}$ satisfying relationship 
$$
\ln(k_{bat}/k_{\beta}) \simeq 3  \eqno{(21)}
$$
effective for an equilibrium scale separating the power-law ('-5/3') and the distributed chaos regions of scales \cite{bb}. It means that with this renormalization we can find the value of $k_{\beta}$ from the first principles.  

\section{Mixing by chaotic motion of the quasi-point vortices}

Besides the Hamiltonian $H$ the Hamiltonian system Eqs. (7-8) on the unbounded plane has additional three independent first integrals (invariants of the motion, see, for instance, Ref. \cite{aref}) :
$$
P_x = \sum_i \Gamma_i x_i,~~~~~~~~~~~ P_y = \sum_i \Gamma_i y_i,  \eqno{(22)}
$$
and 
$$
\mathcal{I}= \sum_i \Gamma_i (x_i^2+y_i^2)    \eqno{(23)}
$$

The integrals Eq. (22) correspond to the two components of the 'linear momentum' ${\bf P}$, and the integral $\mathcal{I}$ Eq. (23) corresponds to 'angular momentum' of the Hamiltonian system Eqs. (7-8). Due to the Noether's theorem \cite{js} conservation of ${\bf P}$ is related to the spatial translational symmetry (spatial homogeneity) and conservation of $\mathcal{I}$ is related to the rotational symmetry (isotropy) \cite{aref}.  \\

\begin{figure} \vspace{-1.2cm}\centering
\epsfig{width=.45\textwidth,file=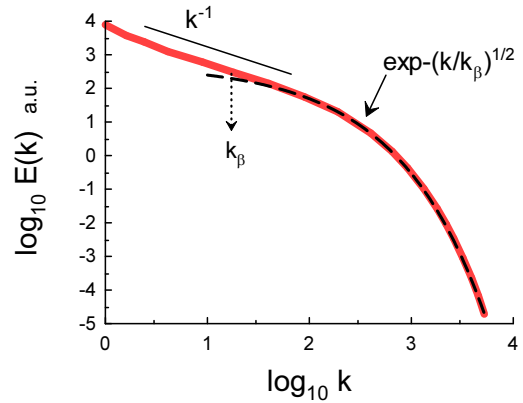} \vspace{-4.1cm}
\caption{Power spectrum of the passive scalar fluctuations produced by chaotic motion of six quasi-point vortices.} 
\end{figure}
\begin{figure} \vspace{-0.5cm}\centering
\epsfig{width=.45\textwidth,file=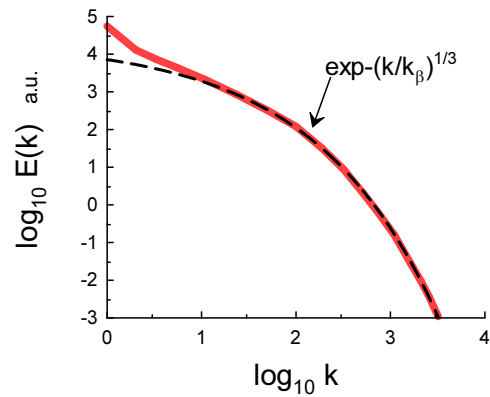} \vspace{-4cm}
\caption{As in the Fig. 5 but for chaotic motion of three quasi-point vortices. } 
\end{figure}
  
  We have already considered a physically meaningful product of invariants in the Eq. (11). Let us consider two additional products of invariants:
$$
I_1 = |{\bf P}|D\Omega   \eqno{(24)}
$$
and
$$
I_2 = |\mathcal{I}|  D\Omega   \eqno{(25)}  
$$
related to the spatial homogeneity and isotropy correspondingly.

  If one of these invariants dominates the distributed chaos, then instead of the relationship Eq. (16) we obtain from the dimensional considerations 
$$
v_c \propto I_1^{1/4}~k_c^{1/4}  \eqno{(26)}
$$  
or
$$
v_c \propto I_2^{1/4}~k_c^{1/2}  \eqno{(27)}
$$   
correspondingly.

  Let us write a general relationship
$$
v_c \propto k_c^{\alpha}  \eqno{(28)}
$$   
Then taking into account the Gaussian distribution of the characteristic velocity $v_c$ we obtain from the Eqs. (28) and (18)
$$
\beta = \frac{2\alpha}{1+2\alpha}   \eqno{(29)}
$$
This relationship results in
$$
E(k) \propto \exp-(k/k_{\beta})^{1/3}  \eqno{(30)}
$$
for the case Eq. (26) (i.e. for the distributed chaos dominated by the spatial homogeneity), or in 
$$
E(k) \propto \exp-(k/k_{\beta})^{1/2}  \eqno{(31)}
$$
for the case Eq. (27) (i.e. for the distributed chaos dominated by the spatial isotropy).\\

  In the paper \cite{yuan} a chaotic motion of six quasi-point vortices was observed in the circular cylinder. The observed leading Lyapunov exponent was positive and approximately equal to 0.7. 
\begin{figure} \vspace{-1.55cm}\centering
\epsfig{width=.45\textwidth,file=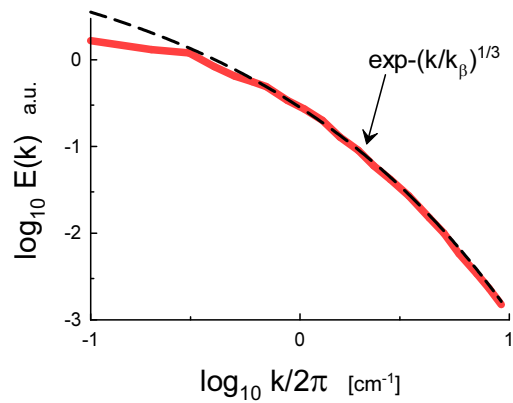} \vspace{-3.5cm}
\caption{Power spectrum of the passive scalar fluctuations in the quasi two-dimensional isotropic homogeneous turbulence} 
\end{figure}
  Fig. 5 shows power spectrum of the passive scalar concentration fluctuations produced by this chaotic motion and corresponding to the initial conditions: ${\bf r}_1(0) = (0.1,20.2)$, ${\bf r}_2 (0) = (0.2,20.8)$, ${\bf r}_3(0) = (0.5,20.5)$,  ${\bf r}_4(0) = (0.1,0.5)$, ${\bf r}_5 (0) = (0.7,0.1),$ and ${\bf r}_6(0) = (0.9,0.2)$, with $\Gamma_1=\Gamma_2=\Gamma_3= -\Gamma_4=-\Gamma_5=-\Gamma_6$=constant. The spectral data were taken from Fig. 13 of the Ref. \cite{yuan}.  The dashed curve in the Fig. 5 indicates correspondence to the spectrum Eq. (31) and the dotted arrow indicates position of the scale $k_{\beta}$. One can see that the distributed chaos was dominated by the rotational symmetry (spatial isotropy) in this case. \\

   Chaotic motion was also observed in the Ref. \cite{yuan} for the three quasi-point vortices for initial conditions: ${\bf r}_1(0) = (0.1,0.7)$, ${\bf r}_2 (0) = (0.2,0.9)$ and ${\bf r}_3(0) = (0.5,0.3)$ with $\Gamma_1=\Gamma_2=\Gamma_3$=constant. The observed leading Lyapunov exponent was positive and approximately equal to 0.3. It often happens for the Hamiltonian systems that for some initial conditions there is a quasi-periodic motion whereas for some other initial conditions there is a chaotic motion.  Fig. 6 shows power spectrum of the passive scalar concentration fluctuations produced by this chaotic motion. The spectral data were taken from Fig. 11 of the Ref. \cite{yuan}. The dashed curve in the Fig. 6 indicates correspondence to the spectrum Eq. (30). It seems that the distributed chaos in this case was dominated by the spatial translational symmetry (spatial homogeneity).

   Since the conservation of the 'angular momentum' of Hamiltonian systems is related by the Noether's theorem to rotational symmetry (isotropy) the 'angular momentum' $\mathcal{I}$ is still a precise invariant for the motion bounded by the circular cylinder, whereas the 'linear momentum' ${\bf P}$ is not, because its conservation is related by the Noether's theorem to spatial translational symmetry (homogeneity) \cite{aref},\cite{js}. Therefore, Fig. 6 indicates that the 'linear momentum' can be still considered as an approximate invariant of the chaotic motion sufficiently far from the boundary and for sufficiently large wavenumbers $k \gg 1/ L$, where $L$ is diameter of the circular cylinder. 
   
\section{Mixing in quasi two-dimensional isotropic homogeneous turbulence}  

  Results of an elaborate laboratory experiment with approximately isotropic and homogeneous quasi two-dimensional turbulence were reported in Ref. \cite{wmg}. The turbulence was studied in a thin layer of fresh water placed above a saline solution (the two layers are separate due to buoyancy). The motion was forced by interaction between a random array of permanent magnets beneath the experimental cell and a steady electric current applied to the saline solution. The secondary flow in the thin (upper) water layer itself was not a magnetohydrodynamic one (see for more detail the Ref. \cite{wmg}). Special measures were taken to minimize vertical velocity components and vertical velocity gradients in the thin upper layer. An irregular array of moving chaotically around the cell vortices was observed. The Reynolds number based on the typical vortex diameter was $Re \sim 500$. A dye solution was used as a passive scalar. The Schmidt number $Sc =2000$ in this experiment. Figure 7 shows power spectrum of the passive scalar fluctuations observed in the experiment (the spectral data were taken from Fig. 17b of the Ref. \cite{wmg}). The dashed curve in the Fig. 7 indicates correspondence to the spectrum Eq. (30) (cf. Fig. 6). 
 
\begin{figure} \vspace{-1.3cm}\centering
\epsfig{width=.45\textwidth,file=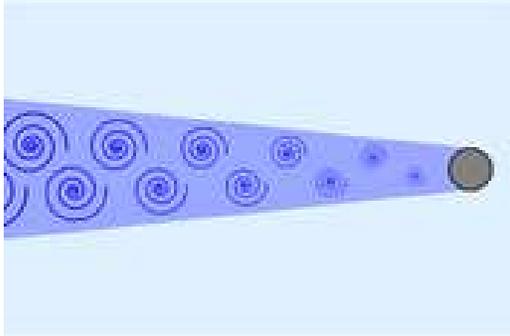} \vspace{-4.2cm}
\caption{Sketch of a vortex street in a flow behind circular cylinder. } 
\end{figure}
\begin{figure} \vspace{-1.2cm}\centering
\epsfig{width=.45\textwidth,file=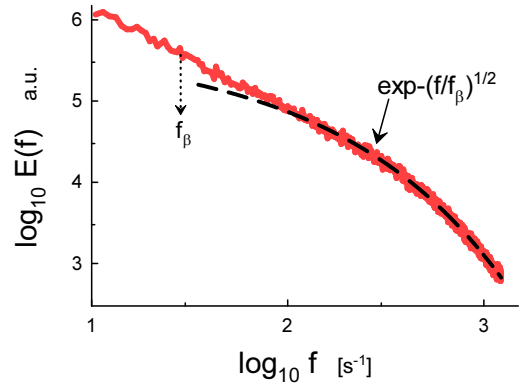} \vspace{-4.25cm}
\caption{Power spectrum of the temperature fluctuations on the wake centerline.} 
\end{figure}
\begin{figure} \vspace{-0.5cm}\centering
\epsfig{width=.45\textwidth,file=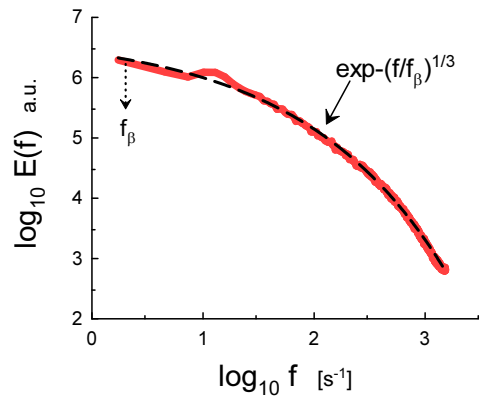} \vspace{-3.5cm}
\caption{Power spectrum of the temperature fluctuations at distance about three cylinder's radii from the the wake centerline. } 
\end{figure}
\section{Wake behind circular cylinder}

  It is interesting to compare results of the previous sections with results of a laboratory experiment with  a passive scalar mixing in turbulent wake behind  a long circular cylinder \cite{kss}. Figure 8 shows a sketch of a so-called vortex street in a flow behind circular cylinder. For turbulent flows the vortex streets are strongly perturbed by the turbulent (chaotic) fluctuations and a good deal of homogenization and isotropization occurs sufficiently far from the cylinder (mainly in the plane orthogonal to axis of the cylinder).\\

  Paper Ref. \cite{kss} reports results of an experiment with a turbulent wake produced by slightly heated circular cylinder in a wind tunnel ($Re_{\lambda} =130$). The temperature fluctuations can be considered a passive scalar in this case.  Figure 9 shows frequency ($f$) power spectrum of the temperature fluctuations on the wake's centerline at distance about 160 cylinder's radii down-flow from the cylinder (cf. Fig. 5). Figure 10 shows power spectrum of the temperature fluctuations at the same distance from the cylinder but at distance about three cylinder's radii from the wake's centerline (cf. Fig. 6). The Taylor hypothesis: $2\pi f = Uk$, where $U$ is the mean constant velocity in the wake, can be applied here \cite{my} for comparison with the Eqs. (30-31) and with Figs. 5-6. The dashed curves in the Figs. 9-10 are drawn for this purpose.

\section{Acknowledgement}

I thank  T. Gotoh, K.R. Sreenivasan and T. Watanabe for sharing 
their data and O.G. Chkhetiani for sending his papers.

\end{document}